# The homogeneous 5D projection and realization of quark and hadron masses

Kai-Wai Wong[1], Gisela A. M. Dreschhoff[1], Högne J. N. Jungner[2]

[1] Department of Physics and Astronomy, University of Kansas, Lawrence, Kansas, USA
[2] Radiocarbon Dating Lab, University of Helsinki, Helsinki, Finland

In this work the homogeneous 5D space-time metric is introduced. Projection operators that map the 5D space-time manifold into a 4D Lorentzian space-time are explicitly given in matrix form. It is emphasized that the concept of proper time is the criterion for the projection. A homogeneous 5D energy-momentum manifold produces naturally the uncertainty principle, and from which we obtained the 5D metric operator, together with the 5D vector and mass-less spinor fields. Hence a coupled product of these two fields is also a solution of the 5D metric operator. Thus the coupling constant is identified as the unit charge. The charged mass-less spinor is coined as the e-trino. Hence the vector field generated by such e-trinos is derived, such that in the 4x1 Hilbert space this vector potential can be identified as the Maxwell monopole potential. Through gauge invariance the concept of charge per unit mass is introduced, which then leads to the mapping of the 5D energy-momentum into that of SU(2)xL and SU(3)xL via the time-shift projection $P_0$ and the conformal space projection $P_1$, respectively. The $P_1$ projection gives us the fractional charged quarks. These quark currents generate both the meson and baryon gluon fields, which in turn generate the meson and baryon masses given in the Eight-Fold-Way representations, removing the necessity of introducing a Higgs vacuum.

## 1. Introduction

The fundamental difficulties with the Gell-Mann standard model for elementary particles [1] is its inability to provide masses for the fractional charged quarks, which can generate the SU(2) and SU(3) representations for the elementary particles. This is because the semi-simple Lie groups can only form a direct product with the Lorentz group, so that no dynamics, that is Lorentz invariant, can ever produce a mass. Because of this mathematical problem, theories such as the Higgs theory were created by introducing a vacuum filled with unobservable bosonic Higgs fields in a condensed state. Apart from that there is no support for such a field, there are several extra difficulties associated with such a theory. First, it has so far proven impossible to generate the gluon fields that are essential to the binding of quarks into the hadrons. Second, such a Higgs theory has not been unified with gravity. Third, the most serious of it comes from the thermal dynamic condensed state that the Higgs fields must remain in. Furthermore, for a condensed state there must exist a critical temperature, meaning that this state can be destroyed. When that happens, all matter will turn to pure energy. This is why any Higgs field must remain in the condensed state up to all known energies that exist in the universe. Recent CERN experiments [2] already make it very unlikely to detect any Higgs field or particle as the data indicated an energy below the 200 GeV level. Therefore, we believe such a theory is fundamentally flawed, and that a more likely reason is that the actual space-time manifold is larger than the Lorentz manifold which contains SU(n) and the Lorentz group L. Furthermore, the realization of mass for the standard model is due to a topological mapping from such a higher space-time manifold into the direct product of SU(n) and L.

In this paper we propose and develop such a theory from a 5D homogeneous space-time manifold. We also show and construct the gluon fields, including calculation of meson and baryon masses, which are very close to experimental values, as well as obtain the bare quark masses, all derived based on preservation of gauge invariance.

## 2. The 5D homogeneous space-time manifold.

In order to achieve what we set out in the Introduction, to find a higher dimensional manifold that would allow us to map into a SU(n)xL, where L is the Lorentz space-time, such that we simultaneously create masses for the quarks given by Gell-Mann's standard model, we should simply look at the 4D metric for a particle with rest mass m:

$$E^2 - \vec{p}^2 = m^2 \qquad (2.1)$$

where the velocity of light c is chosen as 1.

It is obvious that, if we can add a $4^{th}$ momentum component and then fix it equal to m, we regain eq. (2.1). Therefore, the minimum requirement before we go further to prove the required result of finding a mapping such that we recover this larger dimensional manifold into SU(n)xL, is the 5D homogeneous space-time. To proceed further, it is equally obvious that our mapping is really just a projection operation.

We live in the 4D Lorentzian space-time as provided by the metric of Special Relativity:

$$t^2 - \vec{x}^2 = \tau^2 \qquad (2.2)$$

where $\vec{x}$ is the 3D space vector and t the time. In eq. (2.2) we chose as the unit system the velocity of light, c = 1. $\tau$ is the proper time and is a fixed constant such that when we consider the motion of a mass particle it guarantees that the mass velocity can never exceed c. It should be emphasized that this special relativistic restriction does not depend on the exact value of $\tau$, as long as it is finite.

Should we replace this fixed constant $\tau$ by a 4th space dimension variable $x_4$, we have extended the Lorentzian space-time into a homogeneous 5D space-time, where the velocity of anything is always exactly c. This feature of this 5D space-time implies that no massive particle can exist, because if it exists, its energy will be infinite. If we express the 5D space-time metric as follows:

$$(t - x_4 \mid_\tau)(t + x_4 \mid_\tau) = \vec{x}^2 \tag{2.3}$$

or

$$t^2 = (\vec{x} + ix_4 \mid_{\vec{\tau}})(\vec{x} - ix_4 \mid_{\vec{\tau}}) \tag{2.4}$$

where $\vec{\tau}^2 = \tau^2$, the Lorentz metric is recovered when $x_4^2$ in eq. (2.3) and (2.4) is replaced by $\tau^2$. In fact such operations are dimensional projection operations. Projection operators to map this 5D space-time manifold into a 4D Lorentzian space-time are explicitly given in matrix form. Let us define the 5D space-time covariant vector as

$$\vec{\tilde{\xi}} = \begin{bmatrix} it \\ x_1 \\ x_2 \\ x_3 \\ x_4 \end{bmatrix} \tag{2.5}$$

where ~ indicates a 5 component vector.

Then the 5D space-time metric is given by the product of the contra-variant vector $\vec{\tilde{\xi}}^\dagger$ and the covariant vector $\vec{\tilde{\xi}}$. We replace $x_4$ by $\tau$ in eq. (2.3) in order to construct a time-shift 5x4 projection operator $P_0$ to act on the covariant vector

$$P_0 = \begin{bmatrix} 1 & 0 & 0 & 0 & i\tau_0 \dfrac{\partial}{\partial x_4} \\ 0 & 1 & 0 & 0 & 0 \\ 0 & 0 & 1 & 0 & 0 \\ 0 & 0 & 0 & 1 & 0 \end{bmatrix} \tag{2.6}$$

where $\tau_0^2 = \tau^2$, while the time-shift projection operator $P_0^\dagger$ acting on the contra-variant vector is given by

$$P_0^\dagger = \begin{bmatrix} 1 & 0 & 0 & 0 \\ 0 & 1 & 0 & 0 \\ 0 & 0 & 1 & 0 \\ 0 & 0 & 0 & 1 \\ -i\tau_0 \dfrac{\partial}{\partial x_4} & 0 & 0 & 0 \end{bmatrix} \qquad (2.7)$$

To obtain the replacement of $x_4$ by $\tau$ in eq. (2.4) it is given by a conformal space projection

$$P_1 = \begin{bmatrix} 1 & 0 & 0 & 0 & 0 \\ 0 & 1 & 0 & 0 & i\tau_1 \dfrac{\partial}{\partial x_4} \\ 0 & 0 & 1 & 0 & i\tau_2 \dfrac{\partial}{\partial x_4} \\ 0 & 0 & 0 & 1 & i\tau_3 \dfrac{\partial}{\partial x_4} \end{bmatrix} \qquad (2.8)$$

such that $\tau^2 = \sum_{i=1}^{3} \tau_i^2$. We should be aware that projection operators have no inverses. Furthermore, we like to point out that a superposition of $P_0$ and $P_1$ is allowed, provided

$$\tau^2 = \tau_0^2 + \sum_{i=1}^{3} \tau_i^2 \qquad (2.9)$$

These internal degrees of freedom for the resulting Lorentz space-time provide an expansion of choices in the construction of a Riemannian geometry beyond General Relativity [3], where we are allowed only one mass associated to one proper time parameter for deriving the Riemannian curvature.

At this point we would like to point out and discuss a mathematical problem: the Poincaré Conjecture. The conjecture is that all 3 dimension manifolds can be mapped into a sphere. To illustrate the Poincaré problem, we consider only solid volumes in 3D space. Most volumes are mappable into a sphere, except for the doughnut. Recently Perelman [4a,b] showed by the method of Ricci flow, that the doughnut volume is mappable into a sphere plus a chosen axis of orientation. The details of the mathematical proofs are very complicated and are beyond the level of this paper. For those who are interested we refer to Perelman's papers. The result implies mathematically that the three dimension manifold has both symmetric and anti-symmetric representations. In field theory we already know that from the existence of vector and spinor solutions to the Lorentz metric operator. To pictorially explain Perelman's result in a non-rigorous way, let us consider a doughnut shape smoke ring. Depending on the direction of air flow through the doughnut hole, the smoke particles either circulate outward or inward to an observer. These two states are mirror images of each other and cannot be mapped into each other. In another word, they are virtual images to each other, or they are anti-symmetric. Thus the doughnut structure like the spinor field contains two distinct states which cannot be mapped away into a single homogeneous sphere. We can shrink the doughnut hole into a dimensionless line passing through a solid sphere. Further we can displace this line until it reaches the surface of the sphere, at which it becomes a tangent line to the spherical surface. Such a tangent preserves the 2-fold orientations. Before going further with our discussion, we would like to point out that a 5D space-time theory was first introduced by Kaluza and Klein in 1914 [5; 6], which however is very different to our case. Kaluza's 5D space-time is a dimensional extension from the 4D Lorentz space-time. The proper time remains a constant in that 5D space-time and is not fixed by the result of a dimension projection. Therefore none of the theories [7] based on the Kaluza Klein model concern us here.

Apart from the formal mathematics, there are some profound philosophical implications due to the 5D homogeneous space-time metric. By writing it in the form

$$t^2 = \vec{\tilde{x}}^2 \tag{2.10}$$

where $\vec{\tilde{x}}$ is the 4D space vector, we see that when t = 0, there was no space! This interesting solution has the following meaning:

First, there is a beginning of space. Second, time is unidirectional. In another word, there should be chronological order. Thirdly, space is always expanding. Thus if such a 5D space-time contains the universe, then the universe must have a beginning and will grow indefinitely. The existence of a beginning of 5D space-time at t = 0 together with chronology and causality is not the usual view with the arguments relying on the pre-existence of space in which galaxies were created by the Big Bang. In addition the metric (2.10) simply means the measurement of the distance between two space points in 4D space by light. Note, that based only on the space-time metric, there is no reason to have light, not to mention using it for measurement of space.

It is clear that the proportional factor c between time t and the space length does not mean that such a metric implies the existence of a wave field with speed c. However should we make a Fourier transformation to the Lorentz metric we would obtain the 4D Lorentz energy-momentum metric.

$$E^2 - \vec{p}^2 = m^2 \tag{2.11}$$

This metric describes the kinematics of a particle with rest mass m, so that a massive particle can never have a speed exceeding c, the speed of light. Such a particle can be classical. However, it is obvious that if we introduce a 4th component momentum $p_4$, replacing the rest mass m, we extend the energy-momentum metric to the 5D homogeneous energy-momentum space. The significance here is that it can no longer describe the kinematics of a classical massive particle. In fact it must describe a field propagating always with speed c. It is this feature that implies that the 5D homogeneous space-time metric must deal with the kinematics of 'mass-less' fields. Hence events and objects in such a 5D space-time must be given by a probability obtained by the square of the wave function generated by the 5D fields, a concept consistent with quantum mechanics. In order to describe events we need to make measurements. All measurements with such 5D plane waves must have errors irrespective of the average energy and momentum values. By expanding the errors in E, p, and t, x, from the 5D plane wave exponential and keeping the error term of the dot products of the representation by applying the theorem of limit, as it is valid for any E and t values, we get the quantum uncertainty

$$\left|\Delta E \cdot \Delta t\right| = \left|\Delta \vec{\tilde{p}} \cdot \Delta \vec{\tilde{x}}\right| \geq h \qquad (2.12)$$

where $\Delta \vec{\tilde{p}}$ and $\Delta \vec{\tilde{x}}$ are 4D space vectors.

Note that because of the equality between $\left|\Delta E \cdot \Delta t\right|$ and $\left|\Delta \vec{\tilde{p}} \cdot \Delta \vec{\tilde{x}}\right|$, there can only be ONE single Planck's constant 'h'. Therefore, quantum uncertainty is a consequence of the 5D homogeneous space-time metric. We cannot make this argument except for the 5D homogeneous space-time metric operator. Although we have mentioned earlier that the kinematic**s** of a massive particle can be classical based on the Lorentz metric, it however does not exclude from a wave description. In fact, it is perfectly right to describe a massive particle with a de Broglie wave. In order to form a quantum theory we need to first reformulate the 5D energy-momentum metric in the form of a second order differential operator. The 4D Lorentzian operator can then be obtained by the projection operations. Solutions of a differential equation are determined not only by the differential operator, but of equal importance, by the boundary conditions. Hence the quantization of the metric actually provided us with a choice to control the results through the change in boundary conditions, giving us the ability to control our own fate by changing the environment around us, which is philosophically important.

Similar to the 5D space-time case, we introduce the 4th component momentum $p_4$ to extend the 4D Lorentz energy-momentum space into 5D homogeneous energy-momentum space. In fact this is hinted by the famous Einstein equation $E = mc^2$, that mass can be generated from mass-less field energy, by the projection of its momentum onto the energy component: a result independent of the dimension manifold of space. Again following the steps given above, we define the 5 dimension covariant energy-momentum vector, and we obtain the 'energy-shift' and the conformal momentum projection operators:

$$P_0 = \begin{bmatrix} 1 & 0 & 0 & 0 & im\dfrac{\partial}{\partial p_4} \\ 0 & 1 & 0 & 0 & 0 \\ 0 & 0 & 1 & 0 & 0 \\ 0 & 0 & 0 & 1 & 0 \end{bmatrix} \qquad (2.13)$$

and $$P_1 = \begin{bmatrix} 1 & 0 & 0 & 0 & 0 \\ 0 & 1 & 0 & 0 & im_1\dfrac{\partial}{\partial p_4} \\ 0 & 0 & 1 & 0 & im_2\dfrac{\partial}{\partial p_4} \\ 0 & 0 & 0 & 1 & im_3\dfrac{\partial}{\partial p_4} \end{bmatrix} \qquad (2.14)$$

where
$$m^2 = \sum_{i=1}^{3} m_i^2 \qquad (2.15)$$

These two projections produce different Lie group symmetries when the 'massive' charged spinor fields, solutions of the linearized differential operators, are coupled locally to the 'mass-less' vector fields. Thus for the 'massive' spinors they are mutually exclusive. We will discuss in detail the Lie groups involved in the next couple of sections.

To obtain the 5D energy-momentum metric operator, we apply the quantum uncertainty principle, and replace the linear 4 dimension momentum $\vec{p}$ by the differential operator $i\vec{\tilde{\partial}}$ [see Ref. 8] with the unit h = 1, and the energy by the imaginary time t differentiation, we get

$$\sum_{\mu=0}^{4} -\tilde{\partial}^{\mu}\tilde{\partial}_{\mu} \equiv \pentagon \qquad (2.16)$$

where $\mu$ runs from 0 to 4, representing the 5D components. The ~ sign denotes a vector or vector field in the 5D space-time. Because this metric operator is a second order scalar, and is parity and time reversal invariant, the vector fields satisfying it must be symmetric, and therefore are bosonic. It is obvious that the operator (2.16) can be written as the sum of the Maxwellian operator and the second order differential of the 4$^{th}$ coordinate:

$$\pentagon \equiv \square - \frac{\partial^2}{\partial x_4^2} \qquad (2.17)$$

Thus for a 'mass-less' bosonic field satisfying the (2.17) operator, its solution can be expressed in the 4x1 Hilbert space. The portion for the vector fields in the 4D Lorentz space is exactly that of the Maxwell vector potentials. As this representation is valid only if we apply a projection on the 4th component momentum, such that it is evaluated at 0 (zero) for both energy-shift projection $P_0$ as well as for conformal momentum projection $P_1$. The 'mass-less' vector potential fields in the 4D Lorentz space can have sources generated by the charge density distribution and motion of some 'massive' charged particles, such as electrons, which are spinor fields satisfying the linearized Dirac operator. We will discuss these below. Actually the 5D metric operator can be linearized, similar to the Dirac mass-less case, except in 5D. The solutions to such a differential operator is of course a spinor, as we can easily extend the gamma matrix set from the Dirac set of 4 by the addition of a unit matrix to be multiplied to the 5th component of the 5D spinor. Since both such a spinor field and the 5 components vector fields are solutions of the 5D metric operator, their products must also be solutions. The multiplication constant between them is therefore their coupling constant. Should we call this coupling a charge 'e', we can then interpret it such that the mass-less spinor carries such a charge 'e'. In the future we will refer to such a 5D spinor as 'e-trino', to remind our readers that it is also mass-less. Hence in 5D, we can have charged source terms for the 5D vector potentials, similar to the Maxwell potentials in the 4D Lorentz space. On the other hand, for particles without net charge, but having masses, and their fields represented by a scalar potential, such as gravitational potential, the second order derivative operator $-\frac{\partial^2}{\partial x_4^2}$, due to the projection operators $P_0$ and $P_1$, cannot be zero. In fact the mass values in the resulting projected Lorentz space would be x-location and t-time dependent. Thus this resulting space is no longer homogeneous. In another word, we have a Riemannian space. This Riemannian geometry is not the same as that in the Kaluza or Einstein case, since in their case we always keep the proper time constant, such that the 5D space-time is not homogeneous. It is this feature that by choosing a certain order of projections extended over a finite period from the homogeneous 5D space-time we can find the Riemannian curvature at the edge of the homogeneous 5D space-time. With a certain mass distribution form, we can derive the Newtonian Gravitation constant, and create a model for star distribution in a galaxy. Details will be presented elsewhere.

Theodor Kaluza's discovery (1921) [5] that a five dimensional space could combine gravitation and electromagnetism can be based on simply increasing the number of dimensions. Keeping the proper time in the Lorentz metric, one just has to insert an additional coordinate in the metric expression. The result will follow fairly directly when we make this change in the action for the gravitational equations

$$\delta S = \delta \int d^4 x R \sqrt{-g} \tag{2.18}$$

where R is an invariant derived from the curvature tensor, and g is the determinant of the gravitational potentials $g^{\mu\nu}$. These potentials $g^{\mu\nu}$ are also the metric of the line element

$$dS^2 = g^{\mu\nu} dx_\mu dx_\nu \tag{2.19}$$

This action leads to the Einstein gravitational field equations. However, it is also totally different from our 5D homogeneous metric. It is not possible for us to obtain the gravitational field equations without more than the replacement of the 4[th] space variable with only a time-shift projection, which in energy-momentum representation corresponds to the creation of rest mass. How we can overcome this difficulty and obtain the gravitational equation will be demonstrated elsewhere.

The 5D 2[nd] order partial differential operator as given by eq. (2.16) is without a mass term, therefore any field solutions must also be mass-less and only propagates with the speed 'c'. Since this operator (eq. 2.16) is in 5D space-time, it should have a 5 vector potential component set of solutions. If there is no source, then these vector potentials are simply 5D plane waves. If there are any sources for the vector potentials they must also be in the 5D domain and come from the mass-less yet charged spinors. (From now on we shall refer to these mass-less charged spinors as positive and negative e-trinos). According to the first interpretation of the 5D space-time metric the homogeneous 4D space is finite at a given time t, thus it must have an enclosing surface, which is of 3D. Hence the only possibility is a 4D space vector current carried by the charged mass-less spinors within the 4D finite space. As these spinors are solutions of the linearized 5D metric operator, and since these spinors travel with the same speed c in all directions, the net current flow must be normal to the surface enclosing the symmetric 4D volume, unlike the 2D surface enclosing a 3D symmetric, and therefore spherical volume, which has one normal. Such a surface enclosing 4D volume has 3 orthogonal normal axes, according to topology of n dimensional space. Thus we can choose these three axes as 1, 2 and 3, such that the remaining component 4 can be chosen without a current. With such a choice of coordinate representation, we obtain for the 4D space vector current

$$\Box \tilde{A}_\mu = \tilde{j}_\mu \tag{2.20}$$

where $\mu = 0,1,2,3,4$ such that $\tilde{j}_4 = 0$, and $\tilde{j}_0 = e\bar{\tilde{\Psi}}\tilde{\Psi}$.

Note, even though the e-trinos propagate with speed c, they will never reach the 4D space boundary, as the 4D space also expands at the rate c. Hence we can consider this simply as if there is open boundary for the solutions of the vector potential ($\tilde{A}_\mu$).

By solving the 5D vector potentials equation, we obtain

$$\tilde{A}_\mu = \tilde{A}_\mu^0 - \frac{v_\mu e \bar{\tilde{\Psi}}\tilde{\Psi}}{2\pi |\vec{x} - \vec{x}'|} \tag{2.21}$$

where $\bar{\tilde{\Psi}}\tilde{\Psi}$ is the time dependent charge density, and $\mu = 0,1,2,3,4$.

$$\tilde{A}_\mu^0 = c_\mu e^{iEt - i\vec{k}\cdot\vec{x}} \tag{2.22}$$

Apart from the 5D plane wave solutions to all the vector potentials, we have the following additive terms:

$$\tilde{A}_0' = -\frac{e\bar{\tilde{\Psi}}\tilde{\Psi}}{2\pi |\vec{x} - \vec{x}'|} e^{ik_4 x_4'} \tag{2.23}$$

The superscript (') implies it is an additive term to the plane wave solution. This additive term is of course similar to the Coulomb potential in 4D Lorentz space.

While the 3 current components generated by the e-trinos are given by

$$\tilde{A}_i' = -\frac{(\vec{x} - \vec{x}')_i e\bar{\tilde{\Psi}}\tilde{\Psi}}{2\pi |\vec{x} - \vec{x}'|^2} e^{ik_4 x_4'} \tag{2.24}$$

where i = 1,2,3, representing the magnetic monopole potential. The two equations (2.23) and (2.24) can be combined into one form (2.24) by setting $v_0 = 1$. The 3 velocities $v_i$ are restricted to c. The solution of the 5D vector potentials set by eq.(2.24) is in the 4x1 Hilbert space. To reduce these 5D vector potentials to the 4D Maxwell vector potentials, we need to eliminate the 3 components of e-trino currents by applying an energy-shift projection $P_0$ on the 5D charged spinors in the source terms. It is obvious that the spinor will realize a rest mass m, and its 3 remaining momentum components will satisfy special relativity. Therefore its velocity will be less than 'c'. Furthermore, the spinor density given by eq. (2.22) will no longer be a function of the 4th space coordinate variable. It is straight forward to obtain these changes explicitly by applying the projection $P_0$ as given by eq. (2.7) and (2.13) on $\tilde{\Psi}$. Since the source terms after the energy-shift projection exist only in the Lorentz space, the solutions to the 5D vector potentials must be given in the 4x1 Hilbert space, with the portion in the 4D Lorentz space exactly that of the Maxwell potentials. In another word, the $P_0$ projection on the charged spinor has the effect of reducing the 5D vector potentials but preserving a 4x1 direct product given in the 4x1 Hilbert space. Let us start a step by step illustration, as it is illuminating. Notice, the 5D metric operator becomes the Maxwellian operator if and only if the projected 'mass' is exactly zero. Because:

$$P \pentagon P^\dagger = \square + m^2 \qquad (2.25)$$

implies by comparison to

$$P \frac{\partial^2}{\partial x_4^2} P^\dagger = -m^2 \qquad (2.26)$$

where P is either the energy-shift projection $P_0$ or the conformal momentum projection $P_1$. Thus the 4D Lorentz vector potentials equations are given by

$$\square P\tilde{A}_\mu = P \frac{\partial^2}{\partial x_4^2} \tilde{A}_\mu + P\tilde{j}_\mu \qquad (2.27)$$

In order to obtain the exact Maxwell potentials, the projection operators must not contain the $x_4$ variable matrix elements. In other words, $x_4$ is fixed at 0. Hence, the fourth space component vector field equation is decoupled from the rest. Now, turning back to the 5D field equation

$$\square A_\mu = j_\mu \qquad (2.28)$$

where $j_\mu$ must be a combined source term, $\mu = 0,1,2,3,4$.

$$j_\mu = P \frac{\partial^2}{\partial x_4^2} \tilde{A}_\mu + P\tilde{j}_\mu \qquad (2.29)$$

which is certainly not zero.

From eq. (2.24), and ignoring the first term $P \frac{\partial^2}{\partial x_4^2} \tilde{A}_\mu$ in the source as given by eq. (2.29), we obtain

$$\tilde{A}_\mu = \tilde{A}_\mu^0 - \frac{v_\mu e \overline{\Psi}\Psi}{2\pi \left|\vec{x}-\vec{x}'\right|} e^{i\alpha\Phi_0 + ik_4 x_4'} \tag{2.30}$$

where α = e/m, $\Phi_0$ is the flux quanta, (see the next section on gauge invariance) and $\mu$ = 0,1,2,3. Here the current is carried by both the massive Dirac spinors with velocity less than c, and possibly the mass-less charged spinors, which is a plane wave with velocity c. The current carried by these mass-less spinors has the form of a magnetic monopole, which we will discuss later.

$\tilde{A}_\mu^0$ is the plane wave solution as given in eq. (2.22).

The 5$^{th}$ component $\tilde{A}_4$ remains unaffected as a plane wave, and does not exist in the 4D Lorentz space. Note that in our discussion above we have not considered applying the conformal projection $P_1$ to the source terms, as it is clear that with one spinor field the resulting 3 remaining momentum components will become complex, thus the current will also become complex. Since the source terms must remain real, the single spinor $P_1$ projection term is not allowed. We will study this problem later as we study the massive charged spinor solutions. Before we go on to study the spinor solutions let us state that the vector potentials generated by the charged mass-less e-trinos as given by eq. (2.24) can still exist in Lorentz space if the e-trinos are localized by forming Cooper pairs. This possibility can have interesting consequences which we will not discuss in this paper. Furthermore the source term we ignored from eq. (2.29) actually is non-zero as the 5D vector potentials are represented in the 4x1 Hilbert space.

It is 
$$P \frac{\partial^2}{\partial x_4^2} \tilde{A}_\mu = -k_4^2 P \tilde{A}_\mu \tag{2.31}$$

showing that the 5th dimension can affect the electromagnetic fields in the Lorentz space-time domain in which we exist. A view some believe is our ability to communicate with the 5$^{th}$ dimension of the spiritual world. Of course this is true only when $k_4$ is nonzero [9;10]. Actually this term can be cancelled by the current component generated by the mass-less spinors. Thus under normal circumstances, as this component disperses at the rate c to the infinite 3D space, $k_4$ will tend to zero in the final state, and this term will disappear, reducing to only the Maxwellian sources provided by charges.

**3. Proof of SU(2)xL and SU(3)xL due to projections.**

The Dirac spinor field equation is obtained from the linearization of the Klein-Gordon operator through the factorization of the quadratic operator. Following this same method, we can linearize the 5D quadratic operator ⬠. Thus we obtain a 5D spinor field $\tilde{\Psi}$ which satisfies

$$\sum_{\mu=0}^{4} \tilde{\gamma}^{\mu} \tilde{\partial}_{\mu} \tilde{\Psi} = 0 \qquad (3.1)$$

where µ = 0,1,2,3,4, and the 5D gamma matrix $\tilde{\gamma}_{\mu}$ is a 5x5 matrix, such that $\tilde{\gamma}^{\mu}\tilde{P}_{\mu}$ is a scalar. Like the 4D Dirac gamma matrix, this 5D gamma does not contain the 5D variables, and when the projection P is applied it simply reduces to the 4x4 Dirac gamma matrix. Since the 5D vector fields $\tilde{A}_{\mu}$ also satisfy the linearized operator, the product of $\tilde{A}_{\mu}$ and $\tilde{\Psi}$ must also be a solution. Thus we have two possibilities, either they can be coupled together with a plus or minus coupling single value constant 'e', or they are not coupled and are independent of each other. Note the coupling 'e' value must be unique, otherwise the coupled solution is not uniquely defined. The uncoupled solution is allowed if and only if the mass-less spinor is the solution of the reduced Maxwellian 5D manifold in the 4x1 Hilbert space rather than the homogeneous 5D manifold, which we will proof later. When the projection $P_0$ is applied to the charge-coupled 5D spinor and pair it with this uncharged 4D spinor we get the SU(2) leptons, consisting of a charged massive lepton and a neutral mass-less neutrino. Because we have 4D space coordinates, there should be 4 leptons, but remember that the mass-less e-trino is also an allowed state in the reduced Maxwellian space-time, therefore we should have only three massive leptons remaining produced by $P_0$. Trying to find the missing 4$^{th}$ lepton by going to higher energies would be fruitless if the homogeneous 5D space-time theory is correct. Hence we can summarize the 5D spinor solutions into two states, one with a charge 'e', meaning it is coupled to the 5D vector fields, with the coupling constant plus or minus 'e', and one charge-less, belonging to the lower 4 dimension neutrino state. Irrespective of charge because these 5D and 4D spinors have no mass, their velocity is always c. When $P_0$ is applied we get three pairs of lepton and neutrino as experimentally confirmed. However, the pair consisting of a mass-less charged e-trino and its neutrino cannot be confirmed in the 4D Lorentz space-time, and is therefore unobservable. This feature is due to gauge constraint, which will be discussed below.

Now applying the 5D to 4D projection P to eq. (3.1), we obtain

$$\sum_{\mu=0}^{4} (P\tilde{\gamma}^{\mu}\tilde{\partial}_{\mu}P^{\dagger})P\tilde{\Psi} + (P\tilde{\gamma}^{\mu}P^{\dagger})P(\tilde{\partial}_{\mu}\tilde{\Psi}) = 0 \qquad (3.2)$$

Before evaluating the terms in eq. (3.2) we need the following: As mentioned above, for the first 4 terms with µ = 0,1,2,3  $P\tilde{\gamma}^{\mu}\tilde{\partial}_{\mu}P^{\dagger} = \gamma^{\mu}\partial_{\mu}$ , the left hand side gamma is a 5x5 matrix, while the right hand side is the Dirac 4x4 matrix, while the 5th term with µ = 4 gives

$$P_0\tilde{\gamma}^4\tilde{\partial}_4 P_0^{\dagger} = m \qquad \text{and} \qquad P_1\tilde{\gamma}^4\tilde{\partial}_4 P_1^{\dagger} = \sum_{i=1}^{3}\gamma^i m_i \qquad (3.3)$$

Thus under the energy-shift projection $P_0$, the 5 component vector derivative, reduces to the $\gamma^\mu \partial_\mu + m$ 4 component vector derivative, where μ = 0,1,2,3. Hence, the first term in eq. (3.2) for $P_0$ projection simply gives us the Dirac spinor equation, as expected. This Dirac spinor can carry a charge e, or be charge-less, preserving the 5D spinor charge solutions. Thus these Dirac spinors form a SU(2) group. To preserve the invariant factor α = e/m, and gauge invariance is not possible for the mass-less charged spinor, and therefore the e-trino cannot be confined to the Lorentz space-time domain. While for the charge-less spinor, the neutrino, since 'e' goes to zero we need the neutrino mass 'm' to be also zero for α to be an invariant constant. Hence from $P_0$, lepton neutrinos which carry no charge must be absolutely mass-less. Expecting to find a finite mass however small for neutrinos will fail if the 5D theory is correct. An indirect experiment that deduces a mass for neutrinos should be reinvestigated. [Recently the CERN OPERA experiment reported a measurement of the speed of the muon neutrino [11]. It was concluded that its speed was larger than 'c'! Perhaps, the definite conclusion that we can derive from this result is that the muon neutrino must be mass-less. While the larger than c speed, in our opinion, can be the result of multiple possible factors. For example, the massive muon is described by a De Broglie wave packet. A mass-less spinor is a plane wave, with its energy related to its momentum by the factor c. However, there is no restriction to the muon neutrino in the experiment actually being described by a wave packet with a distribution of frequencies $v$. Such a description can be due to a number of factors, including the boundary conditions of the CERN device, as well as possible coupling between the neutrino spin and the vector field potential generated by the charged muon. A wave packet description will cause $<\vec{v}>$ to be always less than $<v>$, where $<\ >$ means averaged over the wave packet. Since the momentum and energy of the muon neutrino were obtained indirectly from the kinematics of the massive muon, an overestimation of $<\vec{v}>$ will give the appearance of a neutrino speed larger than c. It is our opinion that this is a more probable explanation than the muon neutrino being a Tachyon]. When external electromagnetic field coupling is introduced, the second term in eq. (3.2) must satisfy

$$im \frac{\partial}{\partial p_4}(\tilde{\partial}_\mu \tilde{\Psi}) = -eA_\mu \Psi \qquad (3.4)$$

where we only consider the 4D Maxwell potentials, as the massive spinor only exists in 4D Lorentz space. Solving eq. (3.4), we obtain the relationship between the charged 5D spinor and the 4D Dirac charged spinor $\Psi$.

$$\tilde{\Psi} = e^{-i\alpha p_4 \Phi_0} \Psi \qquad (3.5)$$

where $\Phi_0$ is the unit quantum flux

$$\Phi_0 = \sum_{\mu=0}^{3} \oint A_\mu dx^\mu = \frac{2\pi}{e} \text{ n} \qquad (3.6)$$

with α = e/m is the charge per unit mass. It is important to point out that the factor α must be bounded, thus the projected charge spinor must have a finite mass, however small. Hence the mass-less charged e-trino is always confined to the 5D space-time domain. This result reflects that the coupling of the vector potential $A_\mu$ relates the 4D electron spinor to the 5D e-trino spinor through a simple phase gauge transformation, as $P_0$ fixes $p_4 = m$. Because the flux phase is quantized, it implies that under $P_0$ projection the electron spinor solution is Lorentz invariant and thus observable. However for the mass-less e-trino the factor e/m diverges, and thus cannot produce the gauge constraint. Another important point is that all fundamental massive fields must be in the Lorentz space and are fermions, so that we eliminate the limit α equal to zero by excluding Bose condensation. By extending to other leptons, and the fact that charge-less spinor states, the neutrinos, can exist, because it does not have gauge constraint, therefore it is easy to see that $P_0$ projection gives us the SU(2)xL symmetry only for 3 pairs of massive leptons together with their mass-less neutrino.

Now let us turn to the conformal projection $P_1$. $P_1$ acting on the 5D spinor split it into a 3 component spinor, namely

$$P_1 \tilde{\Psi} = \Psi_i \qquad (3.7)$$

where i = 1,2,3.
Now consider their coupling '$e_i$' to the Maxwell potentials. Following equation (3.5), it can be transformed into a phase gauge factor, where α is $e_i/m_i$ = e/m as the un-projected spinor must be independent of $P_0$ or $P_1$. This means the quark to vector potential coupling $e_i$ must be fractions just like $m_i$. Therefore we have derived the fractional charges of the quarks. Now under the $P_1$ projection, we replace $p_4$ by $s_i$ m and get the quark solutions:

$$\Psi_i = e^{i\alpha m s_i \Phi_0} \Psi \qquad (3.8)$$

Since $s_i \Phi_0$ is a fraction flux quanta, $\Psi_i$ is not Lorentz invariant and observable singly. However, it is clear that multiple products of $\prod_{i>2} \Psi_i$ can be made Lorentz invariant when the sum over all the phases is equal to an integer multiple of the flux quanta. It is this constraint that gives us the quark model [1]. The SU(3) representation of the quark model will be discussed in the next section. Our above proof on the spinors is based on the flux quantum produced by the electromagnetic vector potential. The gluon fields coupling to the elementary particles, like the mesons, can be constructed from products of the electromagnetic potentials coupled to the composite quark-pair meson states. These gluon fields will have to satisfy certain sum rules. Again this will be discussed in the next section.

One of the most challenging problems in physics is formulating a fundamental field theory for mesons and baryons, because these elementary particles are well represented by the SU(3) Lie group, but must also obey Lorentz invariance. Hence the mass splitting within each octet cannot be due to kinematics, as semi-simple compact Lie groups can only form a direct product with the Lorentz group [12]. To obtain the SU(3) representations a fractional charged quark model was advanced by Gell-Mann [1] which easily produces the mesons and baryons and their SU(3) representations. However to create the observed mass splitting of these elementary particle masses the interaction between quarks must be given. Unfortunately a fundamental field theory of these quarks is still lacking. In recent years attempts using the Lagrangian principle to formulate a string theory in an effort to learn the actions among these quarks have become very popular. But such an approach is open ended, as we have unlimited choices possible. The projection model from 5D homogeneous space-time to the Lorentz space-time allows no lee-ways, as the 5D homogenous metric operator only has mass-less 5 component vector and spinor field solutions. If the 5D theory is true then we would be able to obtain the interaction fields within the elementary particles directly from these 5D fields. It was found that indeed the resulting spinor is split into 3 states with $s_1$, $s_2$ and $s_3$ fractional charges. Here we will show that these fractional charges are indeed the quarks. Before solving for the fractional charges, let us review the quadratic sum condition from the $P_1$ conformal projection these fractions must satisfy

$$s_1^2 + s_2^2 + s_3^2 = 1 \qquad (3.9)$$

Obviously these three fractions have both positive and negative values. Hence there are actually three pairs. From now on we shall denote the negative set by $\bar{s}_1, \bar{s}_2, \bar{s}_3$.

Let us consider the meson states. Due to gauge invariance requirement, and the fact that mesons are bosons, the minimum for forming them is a pair of quarks, as quarks are spinors, thus they are fermions. Further taking into account that the meson density is given by the product of its complex conjugate wave function multiplied with the wave function, therefore, there exist only the ($q\bar{q}$). Since each q has 3 choices, such mesons must have 9 total states. If they belong to the SU(3) group then they are given by the (1) + (8) representations, satisfying Feynman's Eight Fold way. SU(3) and the SU(2) representation for the leptons cannot be mixed, which means the energy-shift projection $P_0$, and the conformal momentum projection $P_1$ applied to charges cannot be applied together.

To solve for the fractions $s_i$ let us choose one meson state: $s_1 + \bar{s}_2 = 0$, and a second state $s_1 + \bar{s}_3 = 1$. It is obvious that we will get the solution $s_1 = 2/3$; $s_2 = 2/3$ and $s_3 = -1/3$. These fractions are the exact Gell-Mann quarks. These fractional charges will generate the SU(3) group. To show that, let us refer to the Lie group structure. The general Lie group structural constants $C_{jk}^i$ satisfy

$$C_{jk}^i = -C_{jk}^i \qquad (3.10)$$

and
$$C_{is}^p C_{jk}^s + C_{ks}^p C_{ij}^s + C_{js}^p C_{ki}^s = 0 \qquad (3.11)$$

with the fractional charge $s_i$ identified as $C_{is}^p$, then $\bar{s}_i$ is $C_{ki}^s$. Hence equation (3.11) is given by

$$\begin{aligned} & s_1\bar{s}_2 + s_2\bar{s}_3 + s_3\bar{s}_1 \\ & = (2/3)(-2/3) + (2/3)(1/3) + (-1/3)(-2/3) = 0. \end{aligned} \qquad (3.12)$$

Thus we have proved that the quarks generate the SU(3) group structure.

The baryon states are fermion spinors, therefore they must be products of an odd number of quarks. The lowest allowed gauge invariant states contain 3 quarks (qqq). The ($\overline{qqq}$) states are the complex conjugate states, thus such a pair represents a baryon and its anti-baryon and do not constitute new particles. Since we have 3 choices for each quark component in the structure, there are a total of 3x3x3 = 27 states. Given in the SU(3), it produces the (1) + (8) + (8) + (10) representations. Within these representations, we have baryons with charges -2, -1, 0, +1, +2.

## 4. Mass splitting of hadrons (mesons and baryons), and the bare quark masses.

Due to the fact that elementary particles must satisfy gauge invariance and Lorentz invariance, the binding of the quarks within these elementary particles are by quantum gauge confinement and not by any attractive forces between the quarks. In fact the attractive forces can be relatively weak. Thus in conclusion, the strong nuclear force is not just a force, but rather mainly a quantum confinement. The force fields part that contributes to the binding of the quarks in mesons has been found to be meson-gluons [13], which actually obey certain sum rules. The set of gluon fields coupled to mesons which are composites of quarks with their fractional charges can be constructed from the 5D vector potentials which are generated by the constituents quark current (see eq. (2.24)):

$$\tilde{A}^i_\mu = \frac{s_i e v^i_\mu \bar{\Psi}^i \Psi^i e^{-i\alpha p^i_4 \Phi_0}}{2\pi |\vec{x} - \vec{x}'|} \tag{4.1}$$

where $\Psi^i$ is a function of x', and $v_\mu$ is the 4 covariant velocity component. Thus the vector potentials in the Lorentz space is given by

$$A^i_\mu = P_1 \tilde{A}^i_\mu = i s_i m \frac{\partial}{\partial p^i_4} \tilde{A}^i_\mu$$

$$= \frac{s_i^2 m v^i_\mu e(\alpha \Phi_0) \bar{\Psi}^i \Psi^i e^{-i\alpha s_i m \Phi_0}}{2\pi |\vec{x} - \vec{x}'|} \tag{4.2}$$

The phase factor $\alpha s_i m \Phi_0$ for this vector potential is not in multiples of $2\pi$, hence this 4D vector potential is not gauge and Lorentz invariant. However the meson constituents consist of a quark $s_i$ and an anti-quark $\bar{s}_j$, such that $s_i + \bar{s}_j = 0, \pm 1$. Thus the 4D vector fields generated is a product of the two vector potentials produced by the quark $s_i$, and anti-quark $\bar{s}_j$. Hence the resulting meson-gluon field is given by a second rank tensor potential field:

$$J^{ij}_{\mu\mu'} = \frac{e^2 m^2 v^i_\mu v^j_{\mu'}}{(2\pi)^2} (\alpha \Phi_0)^2 e^{-i\alpha \Phi_0 m(s_i + \bar{s}_j)} (s_i^2 \bar{s}_j^2) \frac{(\bar{\Psi}^i \Psi^i)(\bar{\Psi}^j \Psi^j)}{(\vec{x} - \vec{x}')^2} \tag{4.3}$$

where the state (ij) represents an intermediate meson state located at $\vec{x}'$. The separation between the i, j quarks is ignored.

This gluon potential field falls off as square of the distance, therefore it is relatively short ranged. Since its phase factor is an integer of $2\pi$, it preserves gauge invariance. However, this gluon tensor field has a strength factor $s_i^2 \bar{s}_j^2 \neq 1$, and by itself does not satisfy Lorentz invariance, as for Lorentz invariance and completeness there cannot be residual fractions. None the less, because of this non-Lorentz invariant gluon coupling strength between quarks, mass splitting would occur within the meson SU(3) representations. The charge fractions $s_i$ are 2/3, 2/3, -1/3, hence $|s_i \bar{s}_j|$ has 3 values 1/9, 2/9 and 4/9. There is only 1 choice for 1/9, but 4 choices each for 2/9 and 4/9 respectively. That means that through the gluon fields, the meson masses are renormalized into 4 distinct values, with charge degeneracy maintained. Each mass splitting is between a pair of the meson-gluon coupling strengths, such as between $(1/9)^2$ and $(2/9)^2$ or between $(2/9)^2$ and $(4/9)^2$. Hence if we sum over all the choices in the strength factor of the meson-gluon tensor, we get the sum rule

$$(1/9)^2 + 4(2/9)^2 + 4(4/9)^2 = 1 \qquad (4.4)$$

The mass splitting within each octet representation is due to two different coupling strength meson-gluon fields, not all three. It is not due to the phase factor of the meson-gluon fields as they are all gauge invariant. Thus these fields act like potential wells for the mesons. Hence we need to regroup part of the terms with strengths $(2/9)^2$ and $(4/9)^2$, such that the sum rule remains unchanged.
This sum rule can also be rewritten as

$$(1/9)^2 + 4\{(2/9)^2 + (2/9)^2\} + 4\{(4/9)^2 - (2/9)^2\} = 1 \qquad (4.5)$$

This new arrangement and eq. (4.4) allows for splitting each of the potential terms given by $2(2/9)^2$ and $(4/9)^2 - (2/9)^2$ into doublets, similar to the case between $(1/9)^2$ and $(2/9)^2$, which is what we need for the SU(3) octet representations and to preserve Lorentz invariance. Such a summation over all the choices is called the Meson-Gluon Jet Sum Rule. The 9 choices can be identified as a 3x3 separate sum rules of colors and flavors, or 3+3+3. Each 3 consists of a doublet and a singlet mass state. The same doublet degeneracy also must apply to two of the octet representations, thus the representations split the masses due to the strength pair $(1/9)^2$ and $(2/9)^2$, and the degenerate doublet pair $2(2/9)^2/\sqrt{2}$ and $[(4/9)^2-(2/9)^2]/\sqrt{2}$.

The first octet representation for the meson-gluon field strength is based on choosing the intermediate quark states as those with charges (-1/3), (2/3), and (2/3). Thus the mass splitting within this octet would come from $[(-1/3)(1/3)]^2$ and $[(-1/3)(2/3)]^2$. It is further easy to verify the Gell-Mann - Okubo mass relationship formula.

In order to construct the second octet mass splitting we need to reformulate the quark states into effective charges (-1/3), (1/3), and (2/3). It is obvious that we can obtain the new effective quark with (2/3) charge by the linear composition of the two original quarks with both (2/3) charge each. Thus the result is $[(2/3) + (2/3)]/\sqrt{2}$ state. The original (-1/3) quark state remains unchanged, while the new (1/3) effective quark state is given by $[(2/3) + (-1/3)]$ from either of the original (2/3) charged quarks. Hence it is straight forward to see that the new meson-gluon octet strengths are given by $(1/3)^2[(2/3)^2+(2/3)^2]/\sqrt{2} = 2(2/9)^2/\sqrt{2}$ and $[(2/3)^2+(2/3)^2]/\sqrt{2} \cdot [(2/3)^2-(1/3)^2]/2 = [(4/9)^2-(2/9)^2]/\sqrt{2}$. The division by 2 for the strength factor [4/9-1/9] comes from the fact that this strength is shared equally with either of the two original (2/3) charged quarks. With these new gluon field strengths we obtained the major mass splittings within the second meson octet as presented earlier. This reformulation of the gluon fields in terms of conventional quantum assignments produces the notation of strangeness for the mesons in the octet representations.

It is interesting to look at the meson mass corrections due to these gluon fields. To estimate the meson mass splitting, let us approximate the contribution of the gluon potentials by keeping to only the leading perturbations, under such an approximation we get the Klein-Gordon equation for mesons.

$$(\Box + m^{*2})\Phi_n = -n^2 C^2 \Phi_n \qquad (4.6)$$

where m* is the bare projection rest mass, and C is the lowest energy order perturbation correction, while n is obtained from the field strength integers 1, 4, 16 according to the SU(3) octet representations and $\Phi_n$ is the meson field.

If C is much larger than m*, then the meson masses are roughly scaled by n. The renormalized mass M* is given by

$$M^* = nC\sqrt{1 + m^{*2}/n^2 C^2} \qquad (4.7)$$

In order to obtain this meson solution we must calculate the quark-quark binding. We recall from eq. (4.3) the meson-gluon potential. Because of the gauge invariance, which removes the phase factor, the gluon fields do not split the mass degeneracy for the different charged meson states. The charge correction to the mass is due to the differences in the total bare quark mass of each state. Second, the meson wave function $\Phi^{rs}$ is given by $\Psi_r(\vec{x}')\overline{\Psi}_s(\vec{x}')$, such that $\vec{r} = \vec{x} - \vec{x}'$ is the distance between the meson (rs) to the intermediate state (ij). The index (ij) gives the meson binding from the potential with strength proportional to $\left(s_i^2 \overline{s}_j^2\right)$.

Let us consider the non-relativistic Schrödinger equation for the meson binding:

$$\nabla^2 u + 2\mu^*(E-V)u = 0 \qquad (4.8)$$

where $\mu^*$ is the meson-meson reduced mass, and $u = \Phi^{rs}$ and

$$V = \sum_{\mu\mu'} \frac{e^2 m^2 v_\mu v_{\mu'}}{(2\pi)^2}(\alpha\Phi_0)^2 \frac{s_i^2 s_j^2}{(\vec{x}-\vec{x}')^2}$$

The meson has a center of mass, total bare mass $m^* = (|s_r| + |\overline{s}_s|)m$.

Since V is a central force potential, we can expand the meson wave function in terms of spherical harmonics. Hence we reduce eq. (4.8) to its angular momentum equation

$$u_l'' + 2\mu^* \left\{ E - V - \frac{1}{2\mu^*}\frac{l(l+1)}{r^2} \right\} u_l = 0 \tag{4.9}$$

where $u_l = \sum_m R_l Y_{lm}$, and the radial equation is given by

$$R_l'' + \frac{2}{r} R_l' + 2\mu^* \left[ E - \frac{A_l}{r^2} \right] R_l = 0 \tag{4.10}$$

where

$$\frac{A_l}{r^2} = V + \frac{1}{\mu^*}\frac{l(l+1)}{r^2} \tag{4.11}$$

We will use the WKB method to solve eq. (4.11).
Replacing $R_l = r v_l$, and defining

$$g_l = 2\mu^* \left[ E - \frac{A_l}{r^2} \right] \tag{4.12}$$

We get

$$v_l'' + g_l v_l = 0 \tag{4.13}$$

For the existence of bound states we require $g_l > 0$.
In the WKB approximation method we get $v_l = e^{iy_l}$, and eq. (4.13) reduces to

$$y_l'^2 - i y_l'' = g_l \tag{4.14}$$

If $y_l' \approx \sqrt{g_l}$, then

$$\frac{y_l''}{y_l'^2} = \frac{g_l'}{2 g_l^{3/2}} \tag{4.15}$$

There exists a fair approximate solution to eq. (4.15), if

$$|g_l'| \ll 2 g_l^{2/3} \tag{4.16}$$

Suppose we set

$$y_l' = \sqrt{g_l} + \varepsilon_l \tag{4.17}$$

By ignoring $\varepsilon_l'$ and $\varepsilon_l''$, eq. (4.15) becomes

$$g_l + 2\varepsilon_l \sqrt{g_l} - \frac{i g_l'}{2\sqrt{g_l}} = g_l \tag{4.18}$$

Hence we obtain,

$$\varepsilon_l = \frac{ig_l'}{4g_l} \tag{4.19}$$

Substituting eq. (4.19) into eq. (4.7), we get

$$y_l = \int \left[\sqrt{g_l} + \frac{ig_l'}{4g_l}\right] dr = \int \sqrt{g_l}\, dr + \frac{i}{4}\ln g_l \tag{4.20}$$

Thus

$$v_l = e^{iy_l} \approx \frac{1}{g_l^{1/4}} e^{i\int \sqrt{g_l}\, dr} \tag{4.21}$$

The probability of finding the bound state is determined by $|v_l|^2 \propto \dfrac{1}{\sqrt{g_l}}$ (4.22)

If this eigenstate is perpetual then $g_l$ must be equated to 0.

Suppose that for the meson we have only the ground state being perpetual, and it can be approximated non-relativistically, then we obtain the ground state energy

$$E_0 = \frac{A_0}{r_0^2} = nC = M \tag{4.23}$$

Referring back to the meson-gluon potentials as given by eq. (4.8), we note that it is a repulsive potential. The mass M from eq. (4.23) is not from any gluon binding. In fact it is due to the quantum gauge confinement on the meson quarks. As they must lie on the loop radius $r_0$ around the intermediate quark density at $\vec{x}'$, so that these quarks of the meson generate the unit quantum flux in the presence of a vector potential $A_\mu$ (see Fig.1). Note that $A_0$ is proportional to the gluon field strength $\left(s_i^2 \bar{s}_j^2\right)$, and independent of the reduced mass $\mu^*$. $r_0$ is the unit quantum flux radius around $\vec{x}'$, and it must be the same for all meson-gluon potentials. Now substituting the meson field binding energy into the renormalized Klein-Gordon equation, we obtain the resultant meson eq. (4.6). Thus from eq. (4.6) we can obtain all masses for all mesons from the explicit gluon potentials in the SU(3) representations.

If the lightest meson mass within one octet is of the order of 134 MeV (pion) with n = 1, the coupling factor for the charged pions is still 1, because the gluon field phase factor is independent of the meson charge, while the other heavier masses n = 4 will be of the order of 500 MeV (kion and eta), which is close to observed values. It is mainly the gluon couplings that determine the masses of the mesons and not its final charge value. In other words, the meson mass comes from the gluon potential binding energy, not the rest mass obtained from the 5D to 4D projection. The heavy mesons due to quantum number assignments, as we discussed earlier after factoring out common factors, are mixtures of 4 and 16 or $n = (4+4)/\sqrt{2}$ and $n = (16-4)/\sqrt{2}$. The mixing of the two couplings as shown in eq. (4.7) does not change the sum rule. While there is the factor of $1/\sqrt{2}$, the major part of the meson mass comes from the square of mass eigenvalue as given by eq. (4.6), with the renormalization of two states represented by factor 4 and 16. These meson states have masses equal to 695 MeV and 1043 MeV, respectively. Minor mass modifications for individual mesons are due to electromagnetic mass renormalization between different final charge states. The fact that our results are so close to experimental values for the meson masses is truly amazing. In order to keep the focus on the 5D theory, we omit many specific details here on purpose.

However, it is clear the quark charges within the meson generate a mass correction due to basically two factors. 1. Since each quark has a bare mass fraction, there will be a net bare mass contribution on top of M. 2. Since these quarks also possess fractional charges there is a Coulomb potential between each quark pair. For a meson, we have a quark and an anti-quark. If they are of equal charge sign, then this inter-quark Coulomb potential is repulsive and can never affect the unit flux radius $r_0$, but if they are of opposite signs, then this attractive Coulomb potential might be possible to bind the quark pair, resulting in a ground state radius that might or might not be greater than $r_0$. The ground state energy correction is of the order of eV, and will not modify the final meson mass much. However, if the ground state radius of this quark-antiquark pair is less than $r_0$, the unit quantum flux loop radius, then it is clear that the flux must be renormalized by correcting the resulting $A_\mu$ vector field by including the Coulomb potential, and hence arrive at a new reduced $r_o$, which would in effect increase the mass value M (see Fig.1). We shall come back to this discussion later when we investigate the neutron mass.

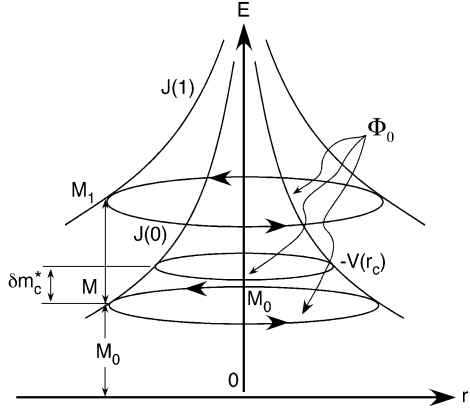

Figure1. The mass gauge diagram depicting the J(i) gluon repulsive potential, resultant mass level splitting $M_0, M_1, \ldots$, with $\delta m_c^*$ resulting from Coulomb correction where $r_c<r_o$. $\Phi_0$, defined by $r_o$, is the quantum unit flux from $A_\mu$. $-V_c(r)$ is the net Coulomb attractive potential with $r_c$ being its ground state orbit. The figure shows that the quarks of the hadron are confined to the $\Phi_0$ loop.

The fields that bind the baryons can be similarly derived from the vector potentials generated by their triplet quark constituents. The result is straight forward. It is a higher 3$^{rd}$ rank tensor potential field, higher than the 2$^{nd}$ rank tensor for the meson-gluons, and its range is even shorter, as it falls off as the cube of distance. Its phase factor like the meson-gluon is gauge invariant, however its strength factor is a fraction, and imaginary. This imaginary factor makes the tensor potentials produce an *im* mass exactly like in the Dirac spinor equation, where there *im* is obtained by the factorization of the Klein-Gordon operator, which we will discuss later in detail. There are 27 choices for the fractions. Note that for example among these 27 states, (1) + (8) + (8) + (10): the state ($q_1q_1q_2$) is equal to $-(q_1q_2q_1)$, therefore they are not different states. Following the construction method of the meson-gluon tensor as given by eq. (4.3), the coupling for the 3$^{rd}$ rank baryon tensor field is proportional to $s_i \cdot s_j \cdot s_k$, such that $s_i + s_j + s_k = n$, where n is an integer $0, \pm 1, \pm 2$. The baryon is the binding of three quarks. The three body is a classical unresolved problem due to the difficulty associated with coordinate transformation so as to handle the interaction potential among the three particles. Therefore our first task is to construct the explicit baryon-gluon potentials generated by the complete set of intermediate baryon states (*ijk*) similar to the gluon potentials for mesons. The coordinates' dependence is an inverse function of $|\vec{x}-\vec{x}_i||\vec{x}-\vec{x}_j||\vec{x}-\vec{x}_k|$. This cyclic product is responsible for the difficulty. To remove this problem we ignore the separations between the 3 quarks and replace them by their center of mass coordinate, and ignore their relative coordinates. With this approximation it is reduced to a form similar to that of the pair mesons interaction via meson-gluon case, and can therefore be solved. In order to evaluate the baryon-gluon binding energy contributions to the baryons, let us do that by analyzing its coupling strength coefficients. There is one choice for coupling 1/27, 6 choices for 2/27, 12 choices for 4/27 and lastly 8 choices for 8/27. Thus we obtain a Lorentz sum rule:

$$(1/27)^2 + 6(2/27)^2 + 12(4/27)^2 + 8(8/27)^2 = 1 \qquad (4.24)$$

Implying baryon binding like meson binding is due to quantum confinement, arising from gauge restraint. The masses of the baryons can be obtained in a similar manner as in the meson case. The tensor field couplings are in proportion of $(1/27)^2$, $(2/27)^2$, $(4/27)^2$, $(8/27)^2$. Thus there would be 4 primarily different masses, like the mesons, only the coupling strength determines their masses. Hence the lightest, the proton and neutron have the same mass. The differences between lightest and heaviest can be as much as 64 times. The heavy baryons are even heavier than many complex nuclei. Separating this sum rule into 3x3x3 separate rules will require introducing one more set of quantum numbers on top of just colors and flavors identified in the meson decays. Let us call the 3 new choices: emotions (happiness, sadness, and indifference). This is certainly worth verification experimentally.

   Previously we demonstrated how the gluon fields generated by the quark-antiquark current can produce the octet meson mass levels using a non-relativistic approximate WKB method, and the results actually provide very close mass levels to those of experimental data. Actually such an approximate method is non-essential, although going through the exercise do give us some physical insights on how the gluon field tensors affect the major mass levels within the SU(3) octet representations. In fact our guiding point in choosing the gluon couplings to the quark currents were such that we can reproduce the experimental values. Here we will discuss the mass levels within the octet and decuplet and show that they are actually the results of a gauge transformation, which we shall coin as 'mass gauge' and that it is totally relativistic invariant. To make our presentation simple and easy to understand, we will first follow the earlier WKB approach, before showing that a unique fully relativistic invariant gauge transformation is derivable which gives the exact experimental data.

The baryons form the 1+8+8+10 representations. The only stable particles are the proton and neutron, and both belong to one octet. As the baryons are made from the product of 3 quarks, with charges (-1/3)e, (2/3)e and (2/3)e, [we shall use each bracket ( ) to represent a specific quark]. Therefore we expect the basic gluon field to have the strength factor (-1/3)$^2$(2/3)$^2$(2/3)$^2$ + (-1/3)$^4$(2/3)$^2$+(-1/3)$^6$, corresponding to the gluon potentials generated by the effective intermediate charges of 1, 0, -1. The term (-1/3)$^2$(2/3)$^2$(2/3)$^2$ represents the intermediate baryon state composed of the product between the two different (2/3) charge quarks. Since the gluon potential produces the binding mass which is always proportional to the gluon strength as we have shown with the non-relativistic WKB method for calculating the binding of the mesons is equally applicable for the baryons under the assumption that all 3 quarks for a baryon can be approximated as located at one space point. The non-relativistic assumption is non-essential as all we need is the conclusion shown by the WKB method that the gluon binding energy is always proportional to the gluon potential strength factor remains valid also for a fully relativistic solution. Thus this octet's masses are predominantly proportional to the gluons with strength V(0)=(2/3)$^2$(2/3)$^2$(-1/3)$^2$ + (2/3)$^2$(-1/3)$^4$+(-1/3)$^6$. We have only one (2/3)$^2$(-1/3)$^4$ term because the additional two terms are added to the higher two levels, hence it is the lowest possible and therefore will give the only stable ground state. Any other gluon potential will give excited states. After considering the electric splitting this ground state produces the proton and neutron particles. The next two excited levels in the octet are generated by V(1)=V(0) + (-1/3)$^4$(2/3)$^2$ and V(2)=V(0)+ 2(-1/3)$^4$(2/3)$^2$.

The vector potential generated by a quark 'i' with covariant velocity $v_\mu^i$ is given by the following expression (see eq. 4.2),

$$A_\mu^i = is_i m \frac{\partial}{\partial p_4} \tilde{A}_\mu^i$$

$$= \frac{s_i^2 m v_\mu^i e(\alpha \Phi_0)}{2\pi |\vec{x} - \vec{x}'|} \cdot e^{-i\alpha s_i m \Phi_0} \bar{\Psi}^i(\vec{x}') \Psi^i(\vec{x}')$$

$$= f^i(s_i) v_\mu^i e^{-i\alpha s_i m \Phi_0} \cdot \bar{\Psi}^i \Psi^i \tag{4.25}$$

where $f^i = \frac{s_i^2 m e(\alpha \Phi_0)}{2\pi |\vec{x} - \vec{x}'|}$.

The baryon-gluon potential as generated by quarks i, j, k is therefore given by

$$J_{\mu_i \mu_j \mu_k}^{ijk} = f^i f^j f^k v_{\mu_i}^i v_{\mu_j}^j v_{\mu_k}^k e^{-i\alpha(s_i+s_j+s_k)m\Phi_0} \cdot \bar{\Psi}_B(s_i, s_j, s_k) \Psi_B(s_i, s_j, s_k) \tag{4.26}$$

where $\Psi_B(s_i s_j s_k) = \Psi^i \Psi^j \Psi^k$.

For a baryon $s_i + s_j + s_k = n$, where n is an integer, due to gauge invariance requirement $e^{-i\alpha(s_i+s_j+s_k)m\Phi_0} = 1$, and the quarks are quantum gauge confined.

Furthermore since these i, j, k quarks are bounded at point $\vec{x}'$, therefore $v_{\mu_i}^i = v_{\mu_j}^j = v_{\mu_k}^k = v_\mu$.

Hence eq. (4.26) reduces to

$$J_\mu^{ijk} = f^i f^j f^k v_\mu^3 \bar{\Psi}_B(s) \Psi_B(s) \quad (4.27)$$
$$= G_\mu(s) \bar{\Psi}_B(s) \Psi_B(s)$$

where $\Psi_B(s) \equiv \Psi_B(s_i, s_j, s_k)$

$$\text{and } G_\mu(s) = O(s) S_\mu \quad (4.28)$$

such that $O(s) = O(s_i s_j s_k)$,

where we factor out the explicit $s_i, s_j, s_k$ dependent coefficient from $f^i f^j f^k$, hence

$$S_\mu = \{\frac{me(\alpha\Phi_0)}{2\pi|\vec{x}-\vec{x}'|}\}^3 v_\mu^3.$$

As $J_\mu^{ijk}$ is real and positive, this gluon potential is a conformal pseudo vector.

The Dirac equation for $\Psi_B$ is now given by

$$\sum_{\mu=0}^{3} \gamma_\mu [\partial_\mu + iO(s) S_\mu] \Psi_B = 0 \quad (4.29)$$

This real pseudo vector gluon potential can be transformed by a complex phase

$$\Psi_B = e^{-i\sum_{\mu=0}^{3} \oint dx_\mu O(s)[S_\mu - i\gamma^\mu M_0]} \Psi'_B \quad (4.30)$$

Substituting eq. (4.30) into eq. (4.29) reduces eq. (4.29) to

$$\sum_{\mu=0}^{3} \gamma_\mu [\partial_\mu + iO(s) S_\mu - iO(s) S_\mu + O(s)\gamma^\mu M_0] \Psi'_B = 0$$

But since $\gamma_\mu \gamma^\mu = 1$, the baryon state equation becomes

$$[\gamma_\mu \partial_\mu + O(s) M_0] \Psi'_B = 0 \quad (4.31)$$

We emphasize the mass $M_0$ is the eigenvalue of the gluon repulsive potential. The baryon binding is the result of gauge invariance, not from an attractive force field. Therefore the lowest ground state eigenvalue of the baryon mass is given by the lowest irreducible $O(s)$ value. In another word, the mass level for the proton and neutron. With $M_0$=44.5 MeV, all the octet and decuplet levels are readily obtained. Hence these gluon interaction potentials give the mass levels values 934.5 MeV, 1115.0 MeV, and 1293.3 MeV, respectively, giving us the major gluon mass level splitting of 178 MeV. Thus these baryon octet masses will satisfy the Gell Mann-Okubo mass formula even when the electric splittings are included. In fact when the electric net charge splittings of the baryons are included we will be able to account for the proton and neutron masses in agreement with experimental data. We will show from the mass splitting of the pions later and from the baryon-gluon potential mass level of 934.5 MeV, ignoring the Coulomb potential effect, and from the proton mass equal to 938.26 MeV, we can get the the bare quark mass of 52.5 MeV. The neutron mass of at 939.55 MeV shall be explained in detail later. It is important to observe that all the mass numbers are generated by just the single gluon field strength created mass level separation of 178 MeV and not the crude value obtained from the Gell Mann-Okubo formula of roughly 180 MeV, which is obtained from averaging over the electric splittings. Because of the detail agreements with all this octet baryon particles, it is another clear indication that the 5D projection theory is likely correct! However none of the second octet baryons have been confirmed. As for the decuplet masses, the major gluon potentials are the same as those in the octet except the delta particles are in the level given by $D(0)= (-1/3)^6+2(-1/3)^4(2/3)^2+(-1/3)^2(2/3)^2(2/3)^2$, where in the decuplet we have a combined zero charge intermediate state, hence we have $2(-1/3)^4(2/3)^2$. The second mass levels come from $D(1)=(-1/3)^6+2(-1/3)^2(2/3)^4$, $D(2)=D(1)+(-1/3)^4(2/3)^2$ and $D(3)=D(2)+(-1/3)^4(2/3)^2$, respectively. Obviously the electric splittings within each potential energy level in the decuplet are different from the octet case as in the octet we only have baryons with net charges of -1, 0 and 1, while in the decuplet case we also have a baryon with net charge of 2. It is straight forward to compute all the decuplet masses following the same procedure as shown in the octet calculations. Before taking care of the electric splittings, we get from the baryon-gluon potential strengths the decuplet 4 levels, namely 1115.0 MeV, 1293.34 MeV, 1471.74 MeV and 1650.147 MeV. It suffice to point out that because the decuplet levels are heavier than the octets the electric splittings due to the bare quark masses becomes relatively less, for example the electric correction to the Omega mass is only 12 MeV, in fact with the bare quark mass of $(18/3)m$ due to the combination of $(-1/3)(-1/3)(-1/3)$ and $3(-2/3)(-2/3)(1/3)$ the omega mass is 1678 MeV. Therefore the decuplet masses are less sensitive to any error that might be caused by the inaccuracy made on the bare quark mass.

It is interesting to point out that the mass splitting levels within the known octet and decuplet did not include the potential strength term $(2/3)^2(2/3)^4$. This term will produce baryon mass in excess of 2840 MeV, which means that there must at least be another octet of baryons with heavier masses. This not yet found octet can be obtained from the jet sum rule given by eq. (4.24).

Apart from the major mass level splittings within each SU(3) representation, the charge state of each particle also produce mass splitting within each mass level. This is due to two facts. First, there is the total net bare quark mass. Second, because quarks carry different charges, therefore Coulomb potential corrections between a pair of quarks must also be added to the mass splitting correction. The details are quite involved, but we can still make a magnitude estimation to both the bare quark masses, and the Coulomb potential corrections.

The mass levels of each octet representations for both the mesons and baryons are due to the gluon potential wells created by the quark currents. These potential wells increase the meson and baryon mass proportional to the potential well strength, which we have discussed earlier. However, there are also inter-level mass splittings due to the charge and other quantum signatures of the hadrons within such a level. It can be seen easily that there are several inter-quark effects within each hadron that would contribute. First, each hadron is a composite of multi-quarks. Thus they have different bare quark masses. Secondly, since all quarks have charges, therefore the inter-quarks Coulomb potentials must play a roll. Thirdly, due to the potential binding between quarks with opposite sign charges, any narrowing of the quark-quark separation distance can in turn lead to the strengthening of the quantum confinement which is already present from gauge restraint. Hence although the quantum signature mass splitting within each mass level is small the calculation for it is more complex. None-the-less, we can still make some simple analysis. To begin we will first study the mass splitting between the charged pions and the neutral pion. Obviously, Coulomb potentials between quarks are present. However, we can see that for the charged pions, they are repulsive and the same. Thus the Coulomb potential will not split the mass between positive and negative pions. For the neutral pion, the net Coulomb potential must be negative, hence compared to the charged pions, the averaged distance between the quark and antiquark within the meson must be less for the neutral pion than for the charged pions. If anything, this Coulomb effect will tends to increase the neutral pion mass, and reduce the charged pion mass after taking into bare quark mass modifications. As a first approximation, let us ignore the Coulomb effects.

The neutral pion must be the superposition of both the (2/3)m or (1/3)m quark and antiquark pairs. Thus by ignoring the Coulomb correction we get

$$M^{*2} \text{ (charged pion)} = M^2 + m^2 \qquad (4.32)$$

The neutral pion is a mixed state between two opposite charged quarks with equal masses 1/3m or 2/3m. Thus this state has a mean square average mass

$$m'^2 = [(4/9 + 16/9)/4] m^2 = 5/9\, m^2 \qquad (4.33)$$

Hence the neutral pion has a resultant mass

$$M'^{*2} \text{ (neutral pion)} = M^2 + 5/9\, m^2 \qquad (4.34)$$

From eq. (4.25) and (4.26) we obtain the bare quark mass m by substituting experimental values of the pion masses

$$m = 3/2 \{M^{*2} \text{ (charged pion)} - M'^{*2} \text{ (neutral pion)}\}^{1/2} \qquad (4.35)$$
$$= 52.5 \text{ MeV}$$

This bare quark mass of 52.5 MeV. would be slightly heavier when Coulomb effects are added. With m given, it is easy to reverse and obtain the pion mass level value M, from equation (4.32). We get the mass level M=129 MeV, as we expect. The charged pion then has a mass of 139 MeV. The neutral pion has a mass of 134.8 MeV. Minor variations with exact masses are of course due to the neglect of the inter quark potentials.

Now let us turn to the baryon octet. The lightest particle is the proton. It should be noted that the proton is composed of the (uud) quarks. The up-quark u has a charge of (2/3)e, while the down-quark has a charge of (-1/3)e. Hence the net bare quark mass for the proton is (5/3)m. Assuming a static model for these 3 quarks with equal distances from each other, we see that the net Coulomb potential vanishes. If our projection model is valid, then m must be a unique mass value for all hadrons. Thus using the value of 52.5MeV we got from the pion splitting, we get from the proton mass, the proton mass level equal to 934.1 MeV, nearly the same as we discussed earlier. Therefore, it is clear that the bare quark mass value is universal. The neutron is made of (udd). Obviously, the net quark mass is (4/3)m, less than that for the proton. However, the Coulomb potentials do not cancel even if they are separated with equal distance. In fact we would obtain a net attractive potential of -1/3V(r). This net potential will reduce the averaged separation distance between the 3 quarks, which in turn will enhance the existing quantum confinement, resulting in an increase to the bare mass correction. Of course such a mass enhancement must be also proportional to the strength factor 1/3. Suppose we assign h as the mass enhancement due to -V(r), then the bare quark mass for the neutron is given by (4/3)m+1/3h. Substituting this into the mass for the neutron, we get 939.6 MeV when h=92.55 MeV (see Figure 1).

This showed clearly, that the Coulomb effect is much larger in the baryons then in the mesons, simply because baryons are smaller particles than mesons. Irrespective of this feature, the rest quark mass of 52.5 MeV is universal for all hadrons. Actually, for the higher mass level hadrons, the corrections due to both the quark rest mass and the Coulomb potentials reduces, and no reverse mass differences caused by the Coulomb effects occur like in the neutron case. We will not go into the details for all the individual hadron masses.

## 5. Conclusions

Although we have only presented the basic points through the projection operations on the 5D homogeneous space-time manifold we have shown that we not only can get the standard model, where quarks have mass, but the explicit formulation of the gluon fields and through them the actual values for the meson and baryon masses. There remain many more results not part of this paper, but which will be included in the book we intend to publish, which would include among others a reformulation of General Relativity that produces no mathematical gravitational singularities inside galactic masses, a model for creation of galaxies, etc. Our main objective has been accomplished clearly, that it is not necessary to introduce a Higgs vacuum in L space. In fact, through our calculation, the quark masses are only of the order of 50 MeV. If we consider closed loops formed by these quarks, their gamma energies emitted will extend from ~100 MeV to infinity. There are no sharp and preferred (energy) channels. We therefore remain cautious that future CERN experiments will be more convincing in proofing the existence of Higgs fields.